\documentclass[12pt, preprint]{aastex}%
\usepackage{amsmath}
\usepackage{graphicx}
\newcommand{\simpl}{{\sc simpl~}}

\newcommand{\simplb}{{\sc simpl}}

\newcommand{\kerrbbtwob}{{\sc kerrbb2}}

\newcommand{\msun}{\rm M_{\sun}}
\newcommand{\Msun}{\rm M_{\sun}}

\newcommand{\rchinu}{\chi^{2}/\nu}

\newcommand{\fsc}{f_{\rm SC}}
\newcommand{\nh}{N_{\rm H}}

\newcommand{\cm}{\rm cm}

\newcommand{\Risco}{R_{\rm ISCO}}

\newcommand{\spin}{a_{*}}

\newcommand{\rxte}{{\it RXTE~}}
\newcommand{\chandra}{{\it Chandra~}}

\newcommand{\rxteb}{{\it RXTE}}

\shorttitle{The Distance, Inclination, and Spin of H1743--322}
\shortauthors{Steiner, McClintock, \& Reid}

\begin{document}

\title{The Distance, Inclination, and Spin of the Black Hole Microquasar H1743--322}

\author{James F.\ Steiner\altaffilmark{1}, Jeffrey E.\ McClintock\altaffilmark{1}, and Mark J.\ Reid\altaffilmark{1}}

\altaffiltext{1}{Harvard-Smithsonian Center for Astrophysics, 60
  Garden Street, Cambridge, MA 02138.} 
\email{jsteiner@cfa.harvard.edu}

\begin{abstract}

During its 2003 outburst, the black-hole X-ray transient H1743--322
produced two-sided radio and X-ray jets.  Applying a simple and
symmetric kinematic model to the trajectories of these jets, we
determine the source distance, $8.5 \pm 0.8$ kpc, and the inclination
angle of the jets, $75\degr \pm 3\degr$.  Using these values, we
estimate the spin of the black hole by fitting its \rxte\ spectra,
obtained during the 2003 outburst, to a standard relativistic
accretion-disk model.  For its spin, we find $\spin = 0.2 \pm 0.3$
(68\% limits); $-0.3 < \spin < 0.7$ at 90\% confidence.  We rule
strongly against an extreme value of spin: $\spin < 0.92$ at 99.7\%
confidence.  H1743--322 is the third known microquasar (after
A0620--00 and XTE J1550--564) that displays large-scale ballistic jets
and has a moderate value of spin.  Our result, which depends on an
empirical distribution of black hole masses, takes into account all
known sources of measurement error.

\end{abstract}


\section{Introduction}\label{section:Intro}

About 50 stellar-mass black holes have been discovered and about two
dozen of these have been well studied at optical or radio wavelengths
\citep{RM06,Ozel_2010}.  They are all accretion-powered X-ray sources
located in X-ray binary systems.  In each system, the X-ray source is
fueled by gas that feeds from a mass-donor star into the black hole's
accretion disk.  Within a few hundred kilometers of the black hole,
the gas reaches a temperature of $\sim10^7$~K and produces a
luminosity that can approach the Eddington limit
($\sim10^{39}$~erg~s$^{-1}$).  More than 80\% of such sources are
transient, with outbursts lasting a year or so followed by years or
decades of quiescence.  Typically, the host binaries have short
orbital periods ($P\sim1$~day) and are comprised of a low-mass
($\lesssim1~M_{\odot}$) secondary star and a $\sim10M_{\odot}$ black
hole.  Presently, neither the masses nor the orbital period of our
featured system, H1743--322, are known.  Nevertheless, as we now
describe, the wealth of data available for H1743--322 (hereafter
H1743) strongly indicates that it is a typical short-period black hole
transient.

Studies of X-ray spectral and timing data leave little doubt that
H1743 contains a black hole primary \citep{Kalemci_2006, JEM_H1743,
  Motta_2010}, notwithstanding the lack of dynamical evidence.  While
large outbursts of H1743 occurred in 1977, 2003 and 2008, our focus
here is on the major 2003 outburst.  During this 9-month active
period, H1743 was observed 170 times using the PCA and HEXTE detectors
aboard the {\it Rossi X-ray Timing Explorer} ({\it RXTE})
\citep{JEM_H1743}.  The source exhibited two distinct phases of
evolution: (1) During the first three months (when the source was
observed on an almost daily basis) H1743 flared continually and
violently and was in the steep power-law (SPL) or intermediate
(SPL:Hard) state (see \citeauthor{RM06} for discussion of these X-ray
states).  On the 47th day of outburst (MJD 52766), an event of central
importance occurred -- the radio/X-ray jets we model were launched
during an intense power-law flare (discussed below).  (2) During the
next four months, the source was locked in the thermal dominant (TD)
state, and the source intensity decayed smoothly and monotonically.
It is primarily these TD-state data that we use to determine the spin
of H1743.


An important X-ray timing result derived from the 2003 outburst was
the discovery of a pair of quasi-periodic oscillations (QPOs) at
240~Hz and 165~Hz \citep{Homan_2005, Remillard_2006}.  Similar
high-frequency (HF) QPOs with a commensurate frequency ratio of 3:2 are
seen for three other dynamically-confirmed black holes (XTE
J1550--564, GRO J1655--40 and GRS 1915+105).

About one year after the onset of the 2003 outburst, bipolar X-ray
jets were discovered and observed a total of three times using
\chandra\ \citep{Corbel_2005}.  Radio observations, which commenced
several months before the X-ray observations, resulted in four
detections of the eastern jet (only), followed about two months later
by a single detection of the western jet \citep{Corbel_2005}.
Large-scale X-ray jets are rare, having been previously observed for
only one other microquasar, namely XTE J1550--564, which is similar in
many respects to H1743 (for comparisons, see \citealt{JEM_H1743}).  In
its 1998 outburst, XTE J1550--564 produced relativistic jets at early
times whose launch date was unambiguously tied to the occurrence of a
remarkable X-ray flare (\citealt{Hannikainen_2009,
  Steiner_j1550jets}).

H1743's X-flare on MJD 52766 showed striking similarities to the giant
flare of XTE J1550--564 (see Figure 12 of \citealt{JEM_H1743}).  Of
particular note, both flares occurred during a dip in the X-ray rms
power (0.1--10 Hz), a jump in frequency of the low-frequency QPOs,
onset of the high-frequency QPOs, and the apex of power-law emission
\citep{Sobczak_QPO, Remillard_2006}.  The similar character of these
two flares, and the coincidence for XTE J1550--564 between the X-ray
flare and the launch date of the jets, motivated us to search the VLA
archive for additional observations of H1743.  This search was
fruitful, and in Section~\ref{section:data} we report three additional
radio jet detections at early times that link the jets to the 2003
X-ray flare.


To deduce the source distance and jet inclination angle of H1743, we
model the proper-motion data derived from the X-ray and radio
observations.  In doing so, we closely follow our recent study of the
large-scale X-ray/radio jets of XTE J1550--564
\citep{Steiner_j1550jets}, which builds on the pioneering work of
\citet{WDL_2003} and \citet{Hao_Zhang_2009}.  Using a model originally
applied to gamma-ray bursts, we concluded that XTE J1550--564 is
embedded in a pc-scale cavity in which the jets expanded unimpeded
until they impacted the cavity walls and rapidly decelerated.  We
apply this same model to H1743 and obtain constraints on the distance
and jet inclination angle (presumed to be the inclination of the spin
axis; see \citealt{Steiner_j1550jets}).

The evolution of H1743's jets have already been studied by
\citet{Hao_Zhang_2009}; however, our aims differ from theirs.  They
were primarily interested in the environment of the black hole.  While
assuming an earlier launch date for the jets, they adopted the nominal
values of distance and inclination ($D=8$~kpc and $i=73\degr$)
suggested by \citet{Corbel_2005}.  Our attention is focused on
deriving accurate constraints on $D$ and $i$ for H1743, which we use
in turn to constrain the spin of the black hole\footnote{We express
  black hole spin in the customary way as the dimensionless quantity
  $a_* \equiv cJ/GM^2$ with $|a_*| \le 1$, where $M$ and $J$ are
  respectively the black hole mass and angular momentum.}.

We measure the spin of H1743 using the continuum-fitting (CF) method
\citep{McClintock_2011, Zhang_1997}.  In the CF method, one estimates
the inner radius of the accretion disk $R_{\rm in}$, which is
identified with the radius of the innermost stable circular orbit
$R_{\rm ISCO}$.  Knowing both $R_{\rm ISCO}$ and $M$ is equivalent to
knowing the spin parameter $a_*$ because $R_{\rm ISCO}/M$ is a
monotonic function of $a_*$, decreasing from 6 to 1 as the spin
parameter increases from 0 to 1 \citep{Bardeen_1972}\footnote{Using
  $c=G=1$.}.  In the CF method, one determines $R_{\rm ISCO}$ by
modeling the X-ray continuum spectrum of the dominant thermal
component using a fully relativistic model of a thin accretion disk.
The observables are X-ray flux, temperature, distance $D$, inclination
$i$, and mass $M$.  In order to obtain reliable values of $a_*$, it is
essential to select X-ray spectra that have a strong thermal component
and to have accurate estimates of $D$, $i$, and $M$.  For H1743, we use
our jet model to determine the first two parameters, and we constrain 
$M$ using the known distribution of black hole masses for X-ray
transient sources.


\section{Data}\label{section:data}
 
To search for the presence of radio jets near the time of their
expected production (Section~\ref{section:Intro}), we examined high
spatial resolution A-configuration VLA images taken early during
H1743's 2003 outburst (see \citealt{JEM_H1743}).  Calibrated data from
the VLA archive for program AR523 on MJD 52779.4, 52782.4, and 52786.4
were imaged using the Astronomical Image Processing System (AIPS) task
{\sc imagr}. The source was detected at 8.4 and 14.9 GHz, but here we
only use the 14.9 GHz data, which had sufficient angular resolution to
clearly resolve source components.  The synthesized beam was
approximately $0.6\arcsec$ by $0.2\arcsec$ elongated
north-south. Fortunately, the jet position angle is almost exactly
east-west \citep{Corbel_2005}, allowing us to identify components
separated by $\gtrsim0.2\arcsec$.  At all three epochs, the source
displayed a dominant component and a weak component offset towards the
west.  At MJD 52779.4, just 13 days after H1743's X-ray flare, their
separation was $166 \pm 20$ mas.  Later, on MJD 52782.4 and MJD
52786.4 the separations were $256 \pm 20$ mas and $288 \pm 20$ mas,
respectively.

The majority of the jet data considered in our analysis are taken from
Tables~1 and 3 of \citet{Corbel_2005}.  These tables provide
jet-source separation measurements for radio and X-ray observations
which were conducted from 6 months onward following H1743's
jet-launching flare.  The X-ray data consist of three $\sim30$~ks {\it
  Chandra} X-ray observations in which both jets were detected.  In
radio, \citet{Corbel_2005} report on five observations from the
Australian Telescope Compact Array (ATCA).  The eastern jet was
present in each image, but the western jet was detected only in the
final observation.  These X-ray and radio observations were carried
out between MJD 52955 and MJD 53092, when the jet-source separations
were in the range $\sim4\arcsec-7\arcsec$.  The substantially larger
angular separations of the eastern jet indicate that it is approaching
and the western jet is receding.

In determining the spin of H1743, we analyze the full set of
\rxteb~PCU-2 ``standard 2'' data obtained during the 2003 outburst,
with the spectra binned into 170 half-day intervals.  These spectra
have been modeled in detail by \citet{JEM_H1743} and
\citet{Steiner_2009}, and we use the same data reduction procedures
here.  Briefly, all the data are dead-time corrected, background
subtracted, and analyzed with the inclusion of a 1\% systematic
uncertainty \citep{Jahoda_2006}.  We standardize all detector
calibrations to the \citet{Toor_Seward} values for the Crab using a
custom model which adjusts both the overall flux normalization and the
spectral shape (see \citealt{Steiner_lmcx3}).  During the early weeks
of the outburst cycle, \rxteb's pointing was offset by $0.32^{\circ}$
from H1743.  We have corrected the fluxes to the full collimator
transmission by assuming a triangular response with FWHM = $1^{\circ}$
(see \citealt{Steiner_2009}).

\section{The Ballistic Jets: Model and Results}\label{section:jets}

Our jet model, which is based on one developed by \citet{WDL_2003},
was first applied in describing gamma-ray-bursts.  Here, we consider a
pair of symmetric jets, each ejected with an initial kinetic energy
$E_0$ and Lorentz factor $\Gamma_0$.  During their expansion, the jets
decelerate as they sweep up gas in their paths.  Assuming adiabatic
expansion, the evolution of each jet is governed by:
\begin{equation}
E_0 = (\Gamma-1) M_0 c^2 + \sigma(\Gamma_{\rm sh}^2-1) m_{\rm sw} c^2,
\label{eq:energy}
\end{equation}
where $\Gamma$ is the bulk Lorentz factor of the jet, $M_0$ the mass
of the ejecta, $\sigma$ is a numerical factor of order
unity\footnote{$\sigma$ ranges from 0.35 in the ultrarelativistic
  limit to 0.73 in the nonrelativistic limit.  For additional details
  concerning our model, see \citet{Steiner_j1550jets}.}, and
$\Gamma_{\rm sh}$ is the Lorentz factor of randomly accelerated
particles at the shock front.  The entrained mass, $m_{\rm sw}$, is
given by $m_{\rm sw} = \Theta^2 m_{\rm p} n \pi R^3/3 $, where
$\Theta$ is the jet half opening angle, $n$ the gas density, $m_{\rm
  p}$ is the mass of a proton, and $R$ the distance traveled by the
jet.

We evolve Eqn.~\ref{eq:energy} in 2-hour time steps, using the
inclination of the jet axis to the observer's line of sight ($\theta$)
to calculate the projected separation ($\delta$) between each jet and
the central source: $\delta(t^\prime) = R(t) {\rm sin}~\theta/D$.
Here, $t^\prime = t \pm R(t) {\rm cos}~\theta/c$ is the observer's
time, which takes into account for each jet the time delay between
H1743's rest frame and the frame of the observer.

\begin{figure}
{\includegraphics[clip=true, angle=90,width=8.85cm]{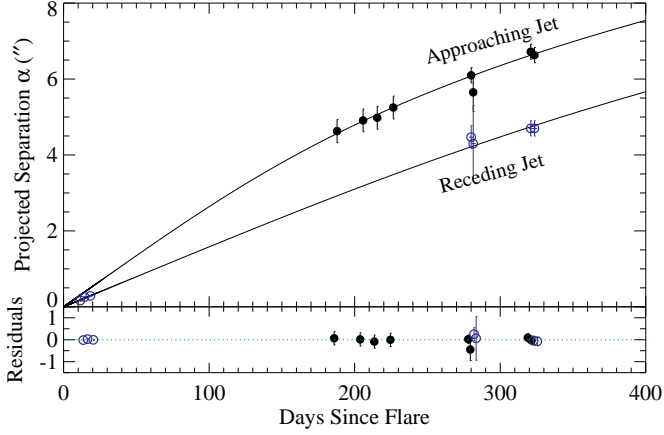}}
\caption{Our best fit model for the motion of H1743's radio and X-ray
  jets.  The eastern jet is marked by filled circles and the western
  jet by open circles.  Fit residuals are shown in the bottom panel
  using a slight offset in time between eastern and western
  jets.}\label{fig:fit}
\end{figure}

\begin{figure}
{\includegraphics[clip=true, angle=90,width=8.85cm]{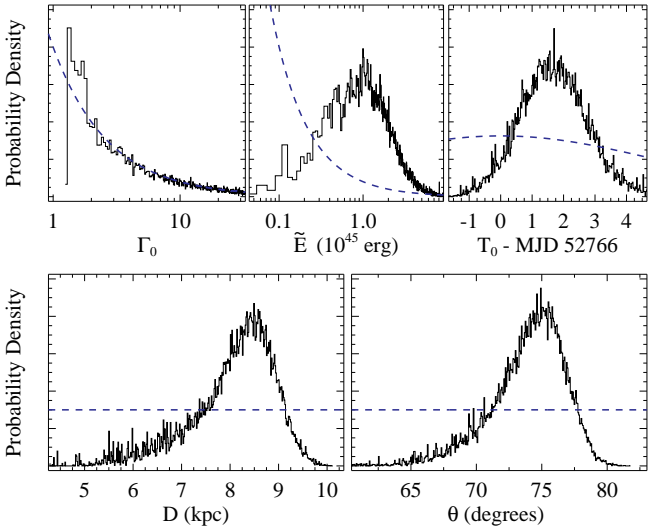}}
\caption{Marginalized probability densities from the MCMC model are
  shown to arbitrary scale.  The prior for each parameter is indicated
  by a dashed line.  $\Gamma_0$ is constrained by its prior at large
  values, but the other parameters show little dependence on their
  priors.  }\label{fig:mcmc}
\end{figure}

Our full model requires just five parameters: $D$, $\theta$,
$\Gamma_0$, the launch date $T_0$, and $\tilde{E}$, the effective
energy\footnote{ $\tilde{E} \equiv E_0 (n/10^{-2} \cm^{-3})^{-1}
(\Theta/1\degr)^{-2}$.  Following our approach for XTE J1550--564
\citep{Steiner_j1550jets}, we scale $\Theta$ and $n$ using typical
values, with density 100 times lower than for the interstellar
medium.}.  Because of the association with the X-ray flare, the prior
on the launch date is taken to be MJD $52766 \pm 5$ days; we adopt a
flat prior on $\theta$, $D$, log($\Gamma_0$), and log($\tilde{E}$).
Our model is fitted via a Markov chain Monte Carlo (MCMC) routine
developed using the Metropolis-Hastings algorithm \citep{MH} which has
been previously applied with this jet model in
\citet{Steiner_j1550jets}.  The chains are evolved until they are well
converged, using $\sim2\times10^5$ elements total\footnote{An
additional $\approx 10^5$ chain elements, which were generated during
training and burn-in phases, were not used.}.

From the VLA data alone, the identification of the pair of radio
sources is ambiguous.  We have applied our model by attributing to the
two radio sources each allowed combination of eastern jet, western
jet, and core.  The most probable interpretation is that the two
sources correspond to emission from the core and the western jet.
Alternative pairings are ruled out at $>97\%$ confidence by our model.

The best fit achieved by the MCMC run is shown in Figure~\ref{fig:fit}
and reaches a goodness of fit $\rchinu = 4.9/9 = 0.54$.  Obviously,
further modification to the model is not needed\footnote{We have
  explored the asymmetric models employed for XTE J1550--564
  \citep{Steiner_j1550jets}; $i$ and $D$ are unchanged.}.
Distributions for the model parameters are shown in
Figure~\ref{fig:mcmc}.  Of chief importance, we find that distance and
inclination are well constrained: $D=8.5 \pm 0.8$~kpc and $i=75\degr
\pm 3\degr$.  This distance places H1743 near the Galactic center,
which is expected, given its projected separation of only $\approx
2\degr$ from the Galactic center.  The time at which the jets were
produced is constrained to $T_0 = $MJD $52767.6 \pm 1.1$ days,
independent of the prior.  This timing supports a connection between
H1743's X-ray flare and the production of its jets.  The speed of the
jets, $\Gamma_0$, has a relatively low maximum a posteriori estimate,
$\Gamma_0 \sim 1.4$, but is poorly constrained at high values and
tracks its prior.  For the kinematic energy of each jet, we obtain a
large uncertainty of $\approx 0.5$ dex centered around $\tilde{E}
\approx 10^{45}$~erg.  This implies that H1743's jets are only about a
tenth as energetic as those produced in the 1998 outburst of XTE
J1550--564 or, alternatively, for H1743 either (1) the density of the
surrounding medium is much lower or (2) the jet opening angle is
substantially smaller.

\section{X-ray Continuum-Fitting Analysis}\label{section:cf}

We now estimate the spin of H1743 by fitting its X-ray spectra.  For
the three crucial input parameters, we use the values of $D$ and $i$
derived in the preceding section and the distribution of black hole
masses discussed below.  All of our analysis is performed using XSPEC
v12.7.0 \citep{XSPEC}.  Following \cite{Steiner_2009} and making minor
adjustments, our spectral model has the form {\sc
  tbabs}(\simplb$\otimes$\kerrbbtwob), where {\sc tbabs} and {\sc
  \kerrbbtwob} are, respectively, the low-energy-absorption and
accretion-disk components.  The component \simpl scatters a fraction
of the thermal disk photons into a Compton power law.  For H1743, this
simple convolution model describes only the broad continuum components
and is unaffected by the inclusion of weaker features due, e.g., to
warm absorbers or spectral reflection.

The four free parameters of the spectral model\footnote{Column density
  $\nh$ is frozen at 2.0$\times 10^{22}~\cm^{-2}$ \citep{Blum_2009}.
  For \kerrbbtwob, limb darkening and returning radiation are switched
  on and the torque at the inner boundary is set to zero.  For {\sc
    simpl}, we use the faster upscattering-only option.} are the (1)
fraction of thermal photons $\fsc$ scattered into the Compton power
law; (2) power-law index $\Gamma$; (3) spin parameter $a_*$; and mass
accretion rate $\dot M$.  Mass, inclination, and distance are varied
in $5\times10^3$ Monte-Carlo samples, and all 170 spectra are fitted
for each setting.  Uncertainty in the absolute calibration of the
X-ray flux is accounted for by randomly varying the overall flux
normalization by 10\% for each triplet setting of $M$, $i$, and $D$
(e.g., \citealt{Steiner_j1550spin_2011}).  We similarly marginalize
over uncertainty in the viscosity parameter $\alpha$ by randomly
assigning either $\alpha = 0.01$ or $\alpha = 0.1$ (e.g.,
\citealt{King_2007, Pessah_2007}), which are two representative values
available to our model.  Both of these uncertainties have a small
effect on $a_*$ compared to our dominant uncertainty, the unknown
black hole mass.  (Uncertainty in the mass accounts for 50\% of our
final uncertainty in $a_*$.)

\begin{figure}
{\includegraphics[clip=true, angle=90,width=8.85cm]{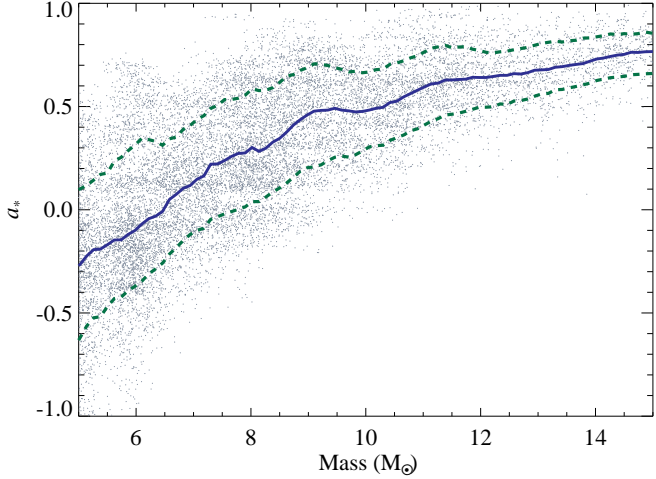}}
\caption{The dependence of spin on black hole mass.  These estimates
  incorporate all sources of measurement error.  The solid line tracks
  the average spin at each mass, and the associated 68\% confidence
  interval corresponds to the region between dashed lines.}
\label{fig:massdep}
\end{figure}

\begin{figure}
{\includegraphics[clip=true, angle=90,width=8.85cm]{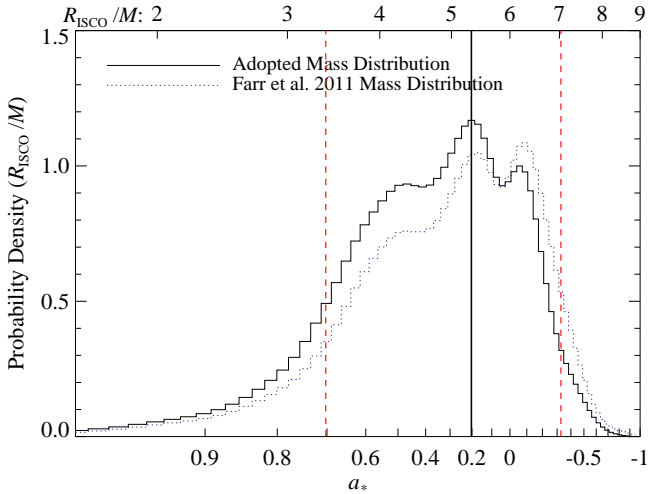}}
\caption{Spin probability for the dimensionless measurement variable
  $\Risco/M$ (top axis) which uniquely determines the spin parameter
  $a_*$ (bottom
  axis).  This spin estimate is obtained by using the adopted
  transient black hole mass distribution from \citet{Ozel_2010} (solid
  curve).  The vertical solid line indicates the maximum likelihood
  spin, while the 90\% confidence range is bounded by dashed vertical
  lines.  For comparison, we also show the spin estimate from using
  the mass distribution of \citet{Farr_2011} (dotted curve).  These
  results take into account uncertainties in $M$, $i$, $D$, $\alpha$
  and the absolute X-ray flux calibration.  For illustration, the
  distributions over $\Risco/M$ have been smoothed using a Gaussian
  kernel with 10\% width.}\label{fig:pdf}
\end{figure}

For each of the $5\times10^3$ parameter settings, we apply our
standard data selection criteria: disk luminosity between 3\% and 30\%
of the Eddington limit; goodness of fit $\rchinu < 2$; and a
power-law normalization $\fsc<25$\% \citep{Steiner_2009}.  Typically,
about 30 spectra pass this screening.  Finally, each of the
$5\times10^3$ samples is given a weight according to the mass
distribution assumed, and random draws are made from the selected
spectra to achieve an estimate of spin.  The dependence between the
inferred value of spin and the black hole's mass is illustrated in
Figure~\ref{fig:massdep}.  As mass is varied from $M = 5\msun$ to
$15\msun$, spin changes from $\spin \approx -0.25$ to $\spin \approx
0.75$.

Recently, \citet{Ozel_2010} compiled all the dynamical measurements of
mass for black hole transients and determined the following best-fit
probability distribution, which we adopt:
\begin{equation}
P(M) =  \left\{
  \begin{array}{l l}
    {\rm Exp}[( 6.30\Msun - M )/1.57\msun ] / 1.57\msun, & \quad M > 6.3\msun,\\
    0, & \quad M \leq 6.3\msun.
  \end{array} \right.
\end{equation}\label{eqn:ozelmass}

In Figure~\ref{fig:pdf}, we show the spin which results when the
\citet{Ozel_2010} mass distribution is assumed.  We find
$\spin=0.20^{+0.34}_{-0.33}$ (68\% confidence interval) with a 90\%
confidence interval of $-0.33<\spin<0.70$.  We also find that extreme
values of spin are ruled out; $\spin < 0.92$ at 99.7\% confidence.
This makes H1743 one of a growing population of black hole
microquasars known to have moderate spin (e.g., A0620--00, $\spin
\approx 0.1$; \citealt{Gou_2010}, XTE J1550--564, $\spin \approx 0.5$;
\citealt{Steiner_j1550spin_2011}).

In Figure~\ref{fig:pdf}, we also show results using the mass
distribution favored by \citet{Farr_2011}.  These authors and
\citet{Ozel_2010} used the same black-hole mass data, but Farr et al.\
found that a power-law distribution gave the best fit, with form $P(M)
\propto M^{-6.4}$ over the mass range $6.1 \msun \leq M \leq 23 \msun$
(and $P(M) = 0 $ elsewhere).  Comparing the Farr et al.\
 distribution with our adopted result, we find that the difference is
minor: $\Delta\spin \approx 0.05$.

\section{Conclusions}

We have modeled the proper motion of the radio and X-ray jets of H1743
that were launched during an X-ray flare.  Based on our purely
kinematic model, we obtain firm estimates of the source distance, $8.5
\pm 0.8$ kpc, and the jet inclination angle, $75\degr \pm 3\degr$.
Using these constraints on $D$ and $i$, we fitted all 170 X-ray
spectra collected during the 2003 outburst of H1743, applied our data
selection criteria, and derived a relationship between spin and black
hole mass.  We then constrained the mass of H1743 using an analytic
distribution for transient systems that are similar to H1743, thereby
arriving at our final result: $\spin = 0.2 \pm 0.3$ ($-0.3 < \spin <
0.7$ at 90\% confidence).  Meanwhile, we rule strongly against an
extreme value of spin: $\spin < 0.92$ at 99.7\% confidence.  Two
similar microquasars have been identified which also produced powerful
jets while harboring black holes with moderate spins: A0620--00
($\spin \approx 0.1 $; \citealt{Gou_2010}) and J1550--564 ($\spin
\approx 0.5$; \citealt{Steiner_j1550spin_2011}).

This is the first successful application of the X-ray
continuum-fitting method that does not rely on any dynamical data to
place constraints on one or more of the input parameters $D$, $M$ and
$i$ -- even the orbital period of H1743 is presently unknown!  Our
constraint on $\spin$ can be tightened once a dynamical estimate of
mass has been obtained.

\acknowledgements

JFS was supported by the Smithsonian Institution Endowment Funds and
JEM acknowledges support from NASA grant NNX11AD08G.  Computations
were performed using the Odyssey cluster which is supported by the FAS
Science Division Research Computing Group at Harvard University.\\

{\it Facilities:} VLA, {\it RXTE}, \chandra

\newcounter{BIBcounter}        
\refstepcounter{BIBcounter}

\end{document}